\documentstyle[12pt]{article}

\newcommand{\wt}{\widetilde}
\newcommand{\wh}{\widehat}
\newcommand{\ha}{{1\over 2}}
\newcommand{\wb}{\bar}
\newcommand{\bmu}{{\bar \mu}}
\newcommand{\bnu}{{\bar \nu}}
\newcommand{\ab}{{(a)}}
\newcommand{\alp}{{(\alpha)}}
\newcommand{\alpp}{{(\alpha')}}
\newcommand{\be}{\begin{equation}}
\newcommand{\ee}{\end{equation}}
\newcommand{\ben}{\begin{eqnarray}\displaystyle}
\newcommand{\een}{\end{eqnarray}}
\newcommand{\refb}[1]{(\ref{#1})}

\newcommand{\V}{{\cal V}}
\newcommand{\sectiono}[1]{\section{#1}\setcounter{equation}{0}}

\begin{document}
{\hfill HUTP-95/A028

\hfill TIFR/TH/95-41

\hfill hep-th/9508064

\vspace {1.0cm}}
\begin{center}{\Large \bf Dual Pairs of Type II String Compactification
}
\end{center}
\bigskip
\bigskip
\centerline{\large Ashoke Sen }
\centerline{\large Tata Institute of Fundamental Research }
\centerline{\large Homi Bhabha Road, Bombay 400005, INDIA }
\medskip
\centerline{and}
\medskip
\centerline{\large Cumrun Vafa}
\centerline{\large Lyman Laboratory of Physics}
\centerline{\large Harvard University }
\centerline{\large Cambridge, MA 02138, USA}

\bigskip
\bigskip

\begin{abstract}
Using a $U$-duality symmetry of type II compactification
on $T^4$ represented by triality action on the $T$-duality group,
and applying the adiabatic argument
we construct dual pairs of type II compactifications in lower
dimensions.
The simplicity of this construction makes it an ideal
set up for testing various conjectures about string dualities.
In some of these models the type II string has
a perturbative non-abelian gauge symmetry.  Examples include
models with $N=2,4,6$ supersymmetry in four dimensions.
There are also self-dual (in the sense of $S-T$ exchange
symmetric) $N=2$ and $N=6$ examples. A generalization of the
adiabatic argument can
be used to construct dual pairs of models with $N=1$ supersymmetry.

\end{abstract}

\vfil\eject

\sectiono{Introduction}
Recently there has been growing evidence for the existence
of many different types of duality symmetries in string
theory\cite{FIQL}-\cite{HaLoSt}.
The existence of duality symmetries has proven to be a powerful
tool in extracting exact results about the vacuum structure
of string compactifications.

One of the best understood duality symmetries is the
conjectured
duality between the type IIA compactification on
$K3$ and heterotic compactification on $T^4$\cite{HuTo1,Vafa}
for which a lot of evidence has recently
emerged\cite{Witten,Sen3,HaSt,VaWi1}.
This duality has itself led to other
dualities in lower dimensions.  In particular upon
further toroidal compactification to 4 dimensions,
this string-string duality leads\cite{Duff,Witten} to the strong/weak
coupling duality of the $N=4$ supersymmetric string
theory in 4 dimensions which had been conjectured
to be a stringy generalization\cite{FIQL}-\cite{GaHa} of the
Olive-Montonen duality\cite{MoOl,Osbo,Sen2,VaWi0,GiPo,Porr}
in $N=4$ Yang-Mills theory. The mechanism for this
reduction is that the $S$-duality of one theory gets mapped
to the $T$-duality for the dual theory. If we assume
that the $T$-duality is not modified quantum mechanically, we are led
to the $S$-duality for the dual theory.  One can also seek
type II--heterotic dualities in theories with lower number of
supersymmetries
in four dimensions and some examples of this have been found
with $N=1,2$ supersymmetry\cite{KaVa,FHSV,VaWi2,HaLoSt}.

As far as string theories with $N=4$ supersymmetry are concerned,
there are
more ways to get them.  On the heterotic side, we can consider
asymmetric orbifolds where, roughly speaking,
the right-movers are compactified
on the six torus and left movers live on an
orbifold\cite{Sche,CHL,ChPo}.
At least some of these models have type II
dual both in six dimensions\cite{ScSe2} and in lower
dimensions\cite{Aspi,VaWi2}.  The
existence of a dual in this
class of models, by the same mechanism mentioned above,
explains the strong/weak duality of the corresponding heterotic theory.
We can get $N=4$ theories in different ways also on the type II side.
We can consider compactifications on manifolds which have $SU(2)$
holonomy but which are not $K3\times T^2$.  We can also
consider compactifications of type II in which all the  4 spacetime
supersymmetries are right-moving, with the left-moving ones
being projected out by the symmetry $(-1)^{F_L}$ (accompanied
by some other symmetries).  In fact these models had
been previously constructed as a way to obtain {\it non-abelian} gauge
theories directly from perturbative
type II strings\cite{BlDoGo,Kaw,Les,FeKo,DiKaVa}.
(We can also consider orientifolding\cite{Orient}
the toroidal type II compactifications
which in many cases is likely to be equivalent
 to the $(-1)^{F_L}$ construction just
discussed \cite{VaWi2}.)

The question arises as to how the $S$-duality of the alternative
$N=4$ models coming from these two type II compactifications can be explained.
This was the motivation for the present paper, and as we will see,
the two alternative $N=4$ type II compactifications will be found to
be dual to one another.  The basic idea is to start from the conjectured
type II $U$-duality in six dimensions\cite{HuTo1} and using the
adiabatic argument of ref.\cite{VaWi2} construct type II dual
pairs with fewer supersymmetries in lower dimensions.

As a by-product we are able to construct many more dual
pairs in d=4, and in particular with $N=1,2$ supersymmetry.
(Construction of dual pairs with $N=1$ supersymmetry requires a
generalization of the adiabatic argument, which will be explained in
section 4.) The construction
of dual type II pairs proceeds in a much simpler way than the corresponding
ones starting from the type IIA-heterotic duality in six dimensions.
The main reason for this is that for the type IIA-heterotic
 duality, given the complexity of the conformal theory
associated with $K3$ compactifications, it is rather difficult
to find the $K3$ conformal theory dual to a given heterotic theory
(and symmetries being modded out are often not geometrical).
For the $T^4$ compactification of type II, which is what
we are considering,
the situation is much simpler and we
can easily translate the geometrical symmetry from one toroidal
compactification to another, with an explicit map.
We can thus produce a large number of dual pairs
and this provides a useful laboratory for testing under what operations
the duality continues to hold.

The organization of this paper is as follows:  In section 2
we review the $U$-duality symmetry of type II string compactified
on $T^4$.
In particular we identify an element $\bar \Omega_0$ of order 2
in this group and show how it conjugates the 8 dimensional
holonomy group (consisting of four left- and four right-mover
holonomies) according to $SO(4,4)$ triality.
In section 3 we discuss how one may use
 adiabatic argument and the type II-type II
six-dimensional duality given by $\bar \Omega_0$ to construct models
with fewer supersymmetry in lower dimensions.  In section
4 we construct a number of such examples, including some
with $N=6,4,2$ supersymmetries in four dimensions.
We also discuss a plausible $N=1$  dual pair, though
it cannot be fully justified by the adiabatic argument.
Among the examples constructed a notable one is
an $N=2$
model in four dimenions which is self-dual (in the sense that
there is an $S-T$ exchange symmetry).

\sectiono{Type II String Compactified on $T^4$}

We begin our discussion in this section by listing the massless fields that
appear in the low energy effective action describing type II string theory
compactified on a four dimensional torus, and the symmetries of this
supergravity theory.  Note that toroidal compactifications
of type IIA and type IIB are equivalent.  For definiteness we
will consider type IIA compactifications.
The scalar fields in this theory are the dilaton $\Phi$,
an $8\times 8$ matrix valued scalar field $M$ parametrizing the coset
$O(4,4)/(O(4)\times O(4))$ corresponding to the moduli associated with
$T^4$, and a set of 8 scalar fields $\psi^\alp$ ($1\le\alpha\le 8$)
associated with the internal components
$A^{[10]}_m$ and $C^{[10]}_{mnp}$ ($6\le m,n,p\le 9$)
of the ten dimensional bosonic fields arising in the
Ramond-Ramond (RR) sector of the theory. $M$ is positive definte, and
satisfies the constraints,
\be \label{e1}
MLM^T=L, \qquad M^T=M\, ,
\ee
where
\be \label{e1a}
L=\pmatrix{-I_4 & \cr & I_4\cr}\, .
\ee
$I_n$ denotes $n\times n$ identity matrix.
The vector fields in this theory consist of U(1) gauge fields $A_\mu^\ab$
($0\le\mu\le 5$, $1\le a\le 8$) arising from the components of the ten
dimensional metric and anti-symmetric tensor fields $G^{[10]}_{m\mu}$ and
$B^{[10]}_{m\mu}$ respectively, and eight U(1) gauge fields $K^\alpp_\mu$
($1\le\alpha'\le 8$) in the RR sector
arising from the components $C^{[10]}_{mn\mu}$, $A^{[10]}_\mu$, and the
dual of $C^{[10]}_{\mu\nu\rho}$. This theory also
has a symmetric tensor field $g_{\mu\nu}$ denoting the canonical Einstein
metric, an antisymmetric tensor field $B_{\mu\nu}$ arising in the
Neveu-Schwarz (NS) sector, and four more anti-symmetric tensor
fields constructed from the components
$C^{[10]}_{m\mu\nu}$. We shall find it convenient to arrange the four
field strengths associated with these anti-symmetric tensor fields into
eight (anti-)self-dual field strengths $D^\alp_{\mu\nu\rho}$
($1\le\alpha\le 8$) satisfying
\be \label{e2}
\wt D^\alp_{\mu\nu\rho} = \big(R_s(ML)\big)_{\alpha\beta}
D^{(\beta)}_{\mu\nu\rho}\, ,
\ee
where $\wt D_{\mu\nu\rho}$ denotes the dual of $D_{\mu\nu\rho}$ in six
dimensions, and for any SO(4,4) matrix $\Omega$,
$R_s(\Omega)$ and $R_c(\Omega)$ denote the two inequivalent spinor
representations of SO(4,4).
We shall also denote by $H_{\mu\nu\rho}$ the field strength
associated with $B_{\mu\nu}$, and by $\wt H_{\mu\nu\rho}$ its dual in
six dimensions.

The action for the resulting $N=4$ supergravity theory in six dimensions
is invariant under an SO(4,4) symmetry generated by an $8\times 8$ matrix
$\Omega$ satisfying
\be \label{e2a}
\Omega L\Omega^T = L\, , \qquad \det\Omega =1 .
\ee
The various fields introduced above transform under this symmetry as,
\ben \label{e3}
M & \to & \Omega M \Omega^T\, , \nonumber \\
\psi^\alp & \to & \big( R_s(\Omega)\big)_{\alpha\beta} \psi^{(\beta)}
\, , \nonumber \\
D^\alp_{\mu\nu\rho} & \to & \big( R_s(\Omega)\big)_{\alpha\beta}
D^{(\beta)}_{\mu\nu\rho} \, , \nonumber \\
A^\ab_\mu & \to & \Omega_{ab} A_\mu^{(b)} \, , \nonumber \\
K^\alpp_{\mu} & \to & \big( R_c(\Omega)\big)_{\alpha'\beta'}
K^{(\beta')}_{\mu} \, . \nonumber \\
\een
$g,H,\wt H$ and $\Phi$
are invariant.
Due to triality, $R_s(\Omega)$
and $R_c(\Omega)$ themselves can be regarded as SO(4,4) matrices. In other
words, they satisfy the relations
\ben \label{e4}
R_s(\Omega) L R_s(\Omega)^T & = & L \nonumber \\
R_c(\Omega) L R_c(\Omega)^T & = & L \, .
\een

In general the maps $R_c(\Omega)$ and $R_s(\Omega)$ are ambiguous, since
they depend on the choice of the basis of the spinor representation. We
shall use this freedom to choose
\be \label{e24}
R_c(\Omega)= R_s^{-1}(\Omega_0 R_s(\Omega) \Omega_0^{-1})\, ,
\ee
where $\Omega_0$ represents the parity transformation matrix
\be \label{em3}
\Omega_0 = \pmatrix{I_2 && \cr & \sigma_3 & \cr && I_4\cr}\, .
\ee
It will be useful for us to study the explicit form of the maps
$R_s(\Omega)$ and $R_c(\Omega)$
in some detail. For this, let us consider the SO(4,4) transformation $\Omega$
to be of the form:
\be \label{e18}
\Omega = \pmatrix{\omega(\theta_L) &&& \cr & \omega(\phi_L) && \cr
&& \omega(\theta_R) & \cr &&& \omega(\phi_R) \cr}\, ,
\ee
where
\be \label{e19}
\omega(\theta) = \pmatrix{\cos\theta & \sin\theta \cr - \sin\theta &
\cos\theta\cr}\, .
\ee
We shall use the shorthand notation $(\theta_L, \phi_L, \theta_R, \phi_R)$
to denote such a matrix. For such an $\Omega$, the map $R_s(\Omega)$
takes the simple form:
\be \label{e21}
R_s(\Omega) = (\theta_L'', \phi_L'', \theta_R'', \phi_R'') \, ,
\ee
where,
\be \label{e22}
\pmatrix{\theta_L''\cr \phi_L'' \cr \theta_R'' \cr \phi_R''}
= A_s \pmatrix{\theta_L \cr \phi_L \cr \theta_R \cr \phi_R \cr}\, ,
\qquad
A_s = \pmatrix{\ha & \ha & \ha & -\ha \cr \ha & \ha & -\ha & \ha \cr
\ha & -\ha & \ha & \ha \cr -\ha & \ha & \ha & \ha}\, .
\ee
Eq.\refb{e24} now gives
\be \label{ebb1}
R_c(\Omega) = (\theta_L', \phi_L', \theta_R', \phi_R') \, ,
\ee
for $\Omega$ of the form \refb{e18}, where,
\be \label{e23}
\pmatrix{\theta_L'\cr \phi_L' \cr \theta_R' \cr \phi_R'}
= A_c \pmatrix{\theta_L \cr \phi_L \cr \theta_R \cr \phi_R \cr}\, , \qquad
A_c = \pmatrix{\ha & -\ha & \ha & -\ha \cr -\ha & \ha & \ha & -\ha \cr
\ha & \ha & \ha & \ha \cr -\ha & -\ha & \ha & \ha}\, .
\ee
{}From eqs.\refb{e18}-\refb{e23} we get
\be \label{ekk1}
R_s(L) = - L\, , \qquad R_s(-L)=L\, , \qquad R_c(L)=L\, .
\ee
Using eqs.\refb{e4} and \refb{ekk1} we see that for a general $\Omega\in
SO(4,4)$ (not necessarily of the form \refb{e18}),
\be \label{em1}
R_s(\Omega^T)= (R_s(\Omega))^T,  \qquad
R_c(\Omega^T)= (R_c(\Omega))^T\, .
\ee
This implies, in particular, that if $\Omega$ is symmetric, then so are
$R_s(\Omega)$ and $R_c(\Omega)$. Also if $\Omega\in O(4)_L\times O(4)_R$
then $R_s(\Omega), R_c(\Omega)\in O(4)_L \times O(4)_R$. These
properties will be useful for us later.

The equations of motion of the theory in fact have a larger symmetry belonging
to the group $SO(5,5)$\cite{CrJu,ScSug,Tanii}.
This group is generated by $5\times 5$ matrices
$\wb\Omega$ satisfying,
\be \label{e6}
\wb\Omega \wb L \wb\Omega^T = \wb L\, , \qquad \det\wb\Omega =1\, ,
\ee
where,
\be \label{e5}
\wb L = \pmatrix{\sigma_1 & \cr & L \cr}\, , \qquad \sigma_1 =
\pmatrix{& 1 \cr 1 & \cr} .
\ee
The SO(4,4) group discussed before is embedded in SO(5,5) as
\be \label{e7}
\wb\Omega = \pmatrix{I_2 & \cr & R_s(\Omega)\cr}\, .
\ee
To see how the full SO(5,5)
symmetry acts on the various fields, it is convenient to
introduce a new matrix valued field
\be \label{e9}
\wb M = \pmatrix{ e^{2\Phi} & -{1\over 2} e^{2\Phi} \psi^T L \psi &
- e^{2\Phi} \psi^T \cr && \cr
-{1\over 2} e^{2\Phi} \psi^T L \psi & e^{-2\Phi} + \psi^T L R_s(M) L \psi &
\psi^T L R_s(M) \cr & +{1\over 4} e^{2\Phi}(\psi^T L \psi)^2 &
+{1\over 2} e^{2\Phi} \psi^T (\psi^T L \psi) \cr \cr
-e^{2\Phi} \psi & R_s(M)L\psi +{1\over 2} e^{2\Phi} \psi(\psi^T L \psi) &
R_s(M) + e^{2\Phi} \psi \psi^T\cr}\, ,
\ee
satisfying,
\be \label{embar}
\wb M \wb L \wb M = \wb L\, , \qquad \wb M^T = \wb M\, .
\ee
We also define,
\be \label{ecov1}
P_\mu =\pmatrix{\vec K_\mu \cr \vec A_\mu} + {\cal O}(\psi)\, ,
\ee
and,
\be \label{ecov2}
Q_{\mu\nu\rho} =\pmatrix{H_{\mu\nu\rho} \cr e^{-2\Phi}
\wt H_{\mu\nu\rho} \cr
\vec D_{\mu\nu\rho}} + {\cal O}(\psi)\, ,
\ee
satisfying the self-duality condition
\be \label{eself2}
\wt Q_{\mu\nu\rho} = \wb M \wb L Q_{\mu\nu\rho}\, .
\ee
${\cal O}(\psi)$ terms in eqs.\refb{ecov1}, \refb{ecov2} reflect the
fact that the field combinations $P_\mu$ and $Q_{\mu\nu\rho}$ which
transform covariantly under SO(5,5) have order $\psi^\alp$ terms in them
which we have not written down explicitly. Since we shall focus on
backgrounds with vanishing $\psi^\alp$, we shall not need to know the
explicit form of these terms.

If $R_S(\wb\Omega)$ denotes the $16\times 16$ matrix
corresponding to
the spinor representation of $SO(5,5)$, then the transformation laws of
various fields under the SO(5,5) transformation is given as follows:
\ben \label{e11}
g_{\mu\nu} & \to & g_{\mu\nu} \, , \nonumber \\
\wb M & \to & \wb \Omega \wb M \wb \Omega^T \, , \nonumber \\
P_\mu & \to & R_S(\wb \Omega)
P_\mu \, , \nonumber \\
Q_{\mu\nu\rho} & \to & \wb \Omega Q_{\mu\nu\rho} \, . \nonumber \\
\een
Using eq.\refb{e7} and the well known decomposition property of the spinor
representation of SO(5,5) under SO(4,4):
\be \label{e14}
R_S(\wb \Omega) = \pmatrix{R_c(\Omega) & \cr & \Omega}\, , \qquad
\hbox{for} \qquad \wb \Omega = \pmatrix{I_2 & \cr & R_s(\Omega)}\, ,
\ee
one can verify that the SO(5,5) transformation laws given in eq.\refb{e11}
are compatible with the SO(4,4) transformation laws given in eq.\refb{e3}.

An SO(4,4;Z) subgroup of SO(4,4) is known to be a symmetry of the type IIA
string theory compactified on $T^4$ order by order in perturbation theory,
and is known as the $T$-duality group of this theory. For definiteness,
we shall always choose our reference $T^4$ to be the one consisting of
four mutually orthogonal cycles with each cycle at the self-dual radius,
and no background $B_{mn}$ field. If $\Gamma_0^{(4,4)}$ denotes the
lattice of integers labelling the momenta and winding numbers on such a
torus, then SO(4,4;Z) is the subgroup of SO(4,4) which leaves this
lattice invariant.

An SO(5,5;Z)
subgroup of the SO(5,5) group has been conjectured to be an exact
non-perturbative symmetry of this string theory\cite{HuTo1}.
A particular element of
this SO(5,5;Z) group, which will play an important role in our
analysis, is
\be \label{e15}
\wb \Omega_0 = \pmatrix{\sigma_1 & \cr & \Omega_0\cr}\, ,
\qquad \Omega_0 = \pmatrix{I_2 && \cr & \sigma_3 & \cr && I_4 \cr}\, .
\ee
We shall conclude this section by listing some of the important properties
of this SO(5,5;Z) transformation $\wb\Omega_0$:

\begin{itemize}

\item{ From eqs.\refb{e7}-\refb{e11} we see that for
$\psi^\alp=0$, $\wb\Omega_0$ induces the transformation
\be \label{e27}
\Phi \to -\Phi \, , \qquad H_{\mu\nu\rho} \to e^{-2\Phi}
\wt H_{\mu\nu\rho}\, .
\ee
Thus the transformation induced by $\wb\Omega_0$ acts on the
$\{\Phi, B_{\mu\nu}\}$ combination in the same way that string-string
duality transformation acts on these fields\cite{Duff,Witten}.}

\item{ $\wb \Omega_0$ conjugates an element of SO(4,4) to another element
of SO(4,4):
\be \label{e16}
\wb \Omega_0 \pmatrix{I_2 & \cr & R_s(\Omega)\cr} \wb \Omega_0^{-1}
= \pmatrix{I_2 & \cr & \Omega_0 R_s(\Omega) \Omega_0^{-1}\cr} =
\pmatrix{I_2 & \cr & R_s(R_c(\Omega))\cr} \, ,
\ee
using eqs.\refb{e24}.
Thus {\it
the net effect of conjugating an} $SO(4,4)$ {\it transformation} $\Omega$
{\it by
the} $SO(5,5)$ {\it transformation} $\wb\Omega_0$ {\it is another}
$ SO(4,4)$ {\it transformation}
$R_c(\Omega)$.
}

\item{ Since conjugation by $\wb\Omega_0$ exchanges the SO(4,4) matrices
$\Omega$ and $R_c(\Omega)$, we see from \refb{e3} that the action of
$\wb\Omega_0$ exchanges the NS sector gauge fields $\{ A_\mu^\ab\}$ with
the RR sector gauge fields $\{ K^\alpp_\mu\}$.
By choosing appropriate basis to describe the set of fields $\{
A_\mu^\ab\}$ and $\{ K^\alpp_\mu\}$ we can have (see eq.\refb{e11}):
\be \label{espin}
R_S(\wb\Omega_0) = \pmatrix{0 & I_8 \cr - I_8 & 0 \cr}\, .
\ee
The origin of the minus sign on the lower left hand corner of this
matrix is the fact that $\wb\Omega_0$ represents a rotation by $\pi$
inside $SO(5,5)$. Thus we expect $R_S(\wb\Omega_0^2)$ to be represented
by $-I_{16}$.
}

\item{ A realization of the duality transformation $\wb\Omega_0$ in
terms of the more familiar duality transformations may be given as
follows. Let us consider the theory as a compactification of the type
IIB theory. This theory has an SL(2,Z) S-duality symmetry in ten
dimensions; let us denote by $\sigma$
the element of this SL(2,Z) group that inverts
the ten dimensional $S$ field. After we compactify the
theory on $T^4$, it also has various $T$-duality symmetries. Let us
denote by $\tau_{mn}$ the duality transformation that takes the
$T$-modulus associated with the $m-n$ plane ($6\le m,n\le 9$) to its
inverse. Then $\wb\Omega_0$ can be regarded as the following combination
of these duality transformations:
\be \label{eduality}
\sigma \cdot \tau_{67} \cdot \tau_{89} \cdot \sigma^{-1} \, .
\ee
}

\item{ The full U-duality group is enhanced if we compactify the theory on one
or more circles, but $\wb\Omega_0$ remains a valid U-duality transformation
in these theories.}

\end{itemize}

\sectiono{Method for Constructing Dual Pairs of Theories}

We shall begin this section by outlining the method for constructing
new pairs of theories that are dual to each other. In view of
potential generalizations to other cases,
 we will
first discuss the method in more generality and then specialize
to the case at hand.  Suppose we consider a theory with $U$-duality
group $G$.  This of course contains the $T$-duality group $H\subset G$
which is the symmetry of the perturbative string theory.  Suppose
 $h,h'\in H$ are elements of finite order $n$
conjugate to one-another in $G$ but
not in $H$, i.e. $h'=g h g^{-1}$ where $g\in G$ but
$g\not \in H$.  By general arguments there is a subspace ${\cal M}
({\cal M}')$
of moduli space of compactifications left invariant by $h(h')$,
which implies that the corresponding perturbative string has $h(h')$
as a symmetry.  Note that $g\cdot {\cal M}={\cal M}'$.  Start
from a string compactification characterized by a point $m\in {\cal M}$.
Compactify on a circle, and mod out the theory by a simultaneous action
of $h$, and a translation of $1/n$ unit along the circle. In other
words, as we go a fraction $1/n$ times around the
circle we identify the internal theory with the action of $h$.
Consider a dual theory starting from $m'=g\cdot m$ on the moduli
space.  Compactify on a circle where as we go a fraction $1/n$ times
around the circle, we come back to the same conformal theory acted
on by $h'$.  By the adiabatic argument of \cite{VaWi2} we
expect the two resulting theories to be dual.  Moreover this
duality cannot be seen perturbatively because $h$ and $h'$
are not conjugate in $H$.  Note that we can compactify
further and mod out by more symmetries to get other dual
pairs in lower dimensions, consistent with the adiabatic argument.

A particularly nice case of the above construction is
when there exists a special element $g\in G$ such that
for all $h\in H$, $ghg^{-1}\in H$, but that $g\not \in H$.
If this happens we can use any finite order $h\in H$ to construct dual pairs.
This luckily is the case for us.

Let us now specialize to our case corresponding to $T^4$ compactification
of type IIA strings
 where $G=SO(5,5;Z)$, $H=SO(4,4;Z)$
and $g={\bar \Omega}_0$ constructed
in the previous section.  So our strategy is as follows.
\begin{itemize}

\item{ We start with type IIA theory compactified on $T^4$ characterized by
the moduli $\langle \wb M \rangle$ which is invariant under some SO(4,4;Z)
transformation $\wh\Omega$.
This theory is equivalent to type IIA
theory characterized by a modular parameter
\be \label{e25}
\langle \wb M' \rangle = \wb \Omega_0 \langle \wb M \rangle \wb \Omega_0^T
\ee
which is invariant under the SO(4,4;Z) transformation
\be \label{el4}
\wh\Omega' = R_c(\wh\Omega)
\ee
according to eq.\refb{e24}.
}

\item{ Now let us further compactify both the theories on $T^2$.
If $\wh\Omega$ is
of order $n$, then we mod out the first theory by simultaneous action of
$\wh\Omega$ and a translation on the first circle of
$T^2$ equal to $1/n$ times
the radius of the circle. Similarly we mod out the second theory by
simultaneous action of the SO(4,4;Z) transformation $\wh\Omega'$ and a
translation on the first circle of $T^2$
equal to $1/n$ times the radius of the
circle. Using the adiabatic argument given in ref.\cite{VaWi2} we expect
the two resulting theories to be equivalent.
}

\end{itemize}

This gives the general method for constructing dual pairs of theories. We
now discuss some of the subtleties and special features
that arise in this construction.

\begin{itemize}

\item{ The adibatic argument by itself does not rule out extra shifts
in the winding direction on the lattice $\Gamma^{(1,1)}$ associated with
the first circle. But in the cases we shall discuss this extra shift is
forced to vanish in order to have level matching between the left and
right moving components of the string states.}

\item{ The construction could be carried out in five dimensions, since the
operations used in modding out the two theories did not involve the
second circle. In fact we can further mod out both the theories by extra
SO(4,4;Z) symmetries accompanied by shifts on the second circle to get more
examples of dual pairs of theories. The adiabatic argument showing the
equivalence of these two theories again applies since we can take the
radius of the second circle to be large.}

\item{ Instead of working with the duality transformations $\wh\Omega$
($\wh\Omega'$) which represent symmetries of $\langle M \rangle$
($\langle M'\rangle$), it is more convenient to work with the
transformations which represent symmetries of the corresponding Narain
lattice.  To do this, let us decompose $\langle M\rangle$ as
\be \label{el1}
\langle M \rangle = \V \V^T\, , \qquad \V \in SO(4,4)\, .
\ee
$\V$ is defined up to an $O(4)\times O(4)$ multiplication from the
right. Invariance of $\langle M\rangle$ under $\wh\Omega$ then implies that
\be \label{el2}
\Omega\equiv \V^{-1} \wh\Omega \V \in O(4) \times O(4)\, .
\ee
Furthermore, the freedom of multiplying $\V$ from the right by an
$O(4)\times O(4)$ matrix translates to a conjugation of $\Omega$ by an
$O(4)\times O(4)$ matrix. Using this freedom we can always bring
$\Omega$ to the form \refb{e18}. This $O(4)\times O(4)$ matrix $\Omega$
now represents the symmetry of the lattice
$\Gamma^{(4,4)}\equiv\V^{-1}\Gamma_0^{(4,4)}$ describing the specific
compactification. ($O(4)\times O(4)$ conjugation of $\Omega$ reflects
the freedom in choosing the axes for describing the lattice.)

Similarly, if we define,
\be \label{el3}
M'=\V'\V^{\prime T}\, , \qquad \Omega' = \V^{\prime -1}\wh\Omega' \V'\in
O(4)\times O(4)\, ,
\ee
then $\Omega'$ corresponds to the symmetry of the Narain lattice
$\Gamma^{(4,4)\prime}\equiv \V^{\prime -1} \Gamma_0^{(4,4)}$ describing
the dual theory. Using eqs.\refb{e25}-\refb{el3} and \refb{e24} we see that
\be \label{el5}
\V' = R_c(\V)\, ,
\ee
and,
\be \label{e26}
\Omega'=R_c(\Omega)\, .
\ee
Modding out the theories described by $\langle M \rangle$ ($\langle M'
\rangle$) by the symmetries $\wh\Omega$ ($\wh \Omega'$) is equivalent to
modding out the corresponding lattices $\Gamma^{(4,4)}$
($\Gamma^{(4,4)\prime}$) by the symmetries $\Omega$
($\Omega'$). Since $\Omega$ is of the form \refb{e18}, $\Omega'$ can
be easily found using eqs.\refb{e21}-\refb{e23}. For a given lattice
$\Gamma^{(4,4)}=\V^{-1}\Gamma^{(4,4)}_0$, the lattice
$\Gamma^{(4,4)\prime}=\V^{\prime -1}\Gamma^{(4,4)}_0$ of the dual
theory can be found using eq.\refb{el5}. We have however not given the
map $R_c(\V)$  explicitly for general $\V$ which is not of the form
\refb{e18}.
}

\item{ Since the relationship between the dilaton and the $B_{\mu\nu}$
fields in the two theories given in eq.\refb{e27}
is identical to the one that describes
string-string duality in six dimensions, the resulting four dimensional
theories will have their $S$- and $T$- moduli exchanged:
\be \label{e28}
S'=T, \qquad T'=S\, .
\ee
}

\item{ For definiteness, let us choose $T^2$ to be the torus with two
orthogonal cycles at self-dual radii, and no background $B_{ij}$ field.
Let $\Gamma_0^{(2,2)}$ denote the lattice associated with this two
dimensional torus. In the basis where the metric is of
the form $\pmatrix{0 & I_2 \cr I_2 & 0}$,
$\Gamma_0^{(2,2)}$ is the lattice of integers.
(A general lattice $\Gamma^{(2,2)}$ can be obtained from
$\Gamma_0^{(2,2)}$ by an $O(2,2)$ boost, which corresponds to switching
on the background $G_{ij}$ and $B_{ij}$ fields.)
In this case the shift on the first circle
accompanying the SO(4,4;Z) transformation $\Omega$ or $\Omega'$ is
represented by a vector
\be \label{e29}
v = \pmatrix{{1\over n} \cr 0 \cr 0 \cr 0 \cr}\, .
\ee
Then the duality symmetry group SO(2,2;Z) associated with $T^2$ is broken
down to a subgroup that preserves the vector $v$ up to lattice translation.
(For N=2 or N=1 supersymmetric theories this duality group is further
modified due to quantum corrections, but for N$\ge$4 supersymmetric
compactification we expect this to be the exact duality symmetry group of
the theory associated with $T^2$.) In order to determine what the surviving
duality group is it is convenient to parametrize an
SO(2,2;Z)=SL(2,Z)$_T\times$SL(2;Z)$_U$ matrix as
\be \label{e30}
\pmatrix{p_1p_2 & p_1q_2 & -q_1q_2 & q_1p_2 \cr
p_1r_2 & p_1s_2 & -q_1s_2 & q_1r_2 \cr -r_1r_2 & -r_1s_2 & s_1s_2 &
-s_1r_2 \cr
r_1p_2 & r_1q_2 & -s_1q_2 & s_1p_2 \cr}
\ee
where,
\be \label{e31}
\pmatrix{p_1 & q_1 \cr r_1 & s_1} \in SL(2,Z)_T, \qquad
\pmatrix{p_2 & q_2 \cr r_2 & s_2} \in SL(2,Z)_U\, .
\ee
{}From eqs.\refb{e29} and \refb{e30} we see that in order to preserve
$v$ up to lattice translations we need,
\ben \label{e31a}
& & p_1p_2 = 1 \, \, \hbox{mod} \, \, n, \qquad
r_1r_2 = 0 \, \, \hbox{mod} \, \, n, \nonumber \\
& & p_1r_2 = 0 \, \, \hbox{mod} \, \, n, \qquad
r_1p_2 = 0 \, \, \hbox{mod} \, \, n.
\een
This gives
\be \label{e32}
\pmatrix{p_1 & q_1 \cr r_1 & s_1} = \pmatrix{1 & q_1 \cr 0 & 1\cr}
\, \, \hbox{mod} \, \, n, \qquad
\pmatrix{p_2 & q_2 \cr r_2 & s_2} = \pmatrix{1 & q_2 \cr 0 & 1\cr}
\, \, \hbox{mod} \, \, n.
\ee
Thus in both the theories SL(2,Z)$_T\times$SL(2,Z)$_U$ is broken down
to $\Gamma_0(n)\times \Gamma_0(n)$. Since the $S$ and $T$ moduli are
exchanged in the two theories, this shows that the S-duality group
SL(2,Z)$_S$ is also broken down to $\Gamma_0(n)$.}

\item{ The full U-duality group of the theory may be found by combining
the subgroup of the six dimensional U-duality group that commutes with
the T-duality transformation $\wh\Omega$ and the full T-duality group of
the four dimensional theory. The element $\wh\Omega$ itself is a
symmetry of the quantum theory\cite{Vafa-q}, but it does not have the
interpretation of a conventional duality symmetry since it leaves all
the moduli fields invariant.
}

\item{ Finally, we note that after modding out the two theories
by appropriate group of transformations, we can switch on the
Wilson lines for the background six dimensional gauge fields that
survive the projection by $\Omega$ ($\Omega'$).  For
NS background fields this will correspond to a deformation of the
internal lattice $\Gamma^{(6,6)}$ away from its factorized from
$\Gamma^{(4,4)}\times \Gamma^{(2,2)}$ consistent with invariance under
$\Omega$ ($\Omega'$). For RR Wilson lines, there is
no such simple geometrical
interpretation. Since the SO(5,5) transformation $\wb\Omega_0$
maps NS sector gauge fields to RR sector gauge fields and vice versa, we
see that a deformation of the lattice $\Gamma^{(6,6)}$ from its
factorized form in one theory corresponds to switching on RR Wilson
lines in the other theory, and vice versa.
}

\end{itemize}

\sectiono{Explicit Examples of Dual Pairs of Models}

In this section we shall construct explicit examples of dual pairs of
models following the procedure outlined in the previous
section.  We will concentrate on constructing supersymmetric
models in four dimensions, though other dimensions can
also be considered.  In type II compactifications, the number
of spacetime supersymmetries we can obtain from each left-
or right-mover Hilbert space is 0,1,2 or 4, corresponding to compactifications
with left- or right-moving holonomy in $SO(6),SU(3),SU(2),1$.
We shall call a theory with $n$ left moving and $m$ right moving
space-time supersymmetry
a theory of the $(n_L,m_R)$ type.
Combining left- and right-moving supersymmetries we can
thus obtain any $N=n+m$ between 0 and 8 except $N=7$ (the $N=7$
presumably gets automatically promoted to $N=8$) --
for some examples
of such models see \cite{BlDoGo,Kaw,Les,FeKo,DiKaVa}.
As far as adiabatic argument is concerned, we need to leave at least
one circle invariant, which implies that we can have all the above
holonomy groups except $SU(3)$.  Thus if we wish to apply the adiabatic
argument we will have to restrict to $0,2,4$ supersymmetries from each
side, which gives us possible
total supersymmetries as $N=2,4,6,8$ (where we omit $N=0$ because
it is not clear how to make sense of
non-supersymmetric string theories).
The $N=8$ theory can be obtained in a unique way, which is continuing
toroidal compactification down to four dimensions. $N=6$ can
be constructed by decomposing $6=4+2$ and so we use toroidal
compactification on one side and $SU(2)$ holonomy on the other side.
$N=4$ theories can be constructed in two ways by decomposing
$4=2+2$ or $4=4+0$.  In the first case we can consider compactifications
with $SU(2)$ holonomy for both left- and right-movers and so
we can use geometric compactifications.  For the latter case we can
use $SO(6)$ holonomy (or $(-1)^{F_L}$) on one side and toroidal
compactification on the other.  For $N=2$ theories, again there
is a unique way of decomposition as $2=2+0$ corresponding to
$SU(2)$ holonomy on one side and $SO(6)$ holonomy on the other.

Note that specifying the holonomy we are modding out by
does not uniquely specify the action.  There are two
potential other choices: 1- There may be inequivalent
realizations of the same holonomy group action on different
lattices. 2-We can also accompany the holonomy action by
translations in the internal theory, which when
using the duality map discussed in section 2 get mapped
to turning on RR Wilson lines in the dual theory.  Both
of these considerations will be relevant below in connection
with constructing the dual to $N=4$ models of ref.\cite{DiKaVa}.

It turns out that the simplest models one can construct
with $N=2$ and $N=6$ are self-dual, mirroring the fact
that there is a unique way of decomposing the supersymmetries
between the left- and right-movers. On the other hand,
the two classes of $N=4$
models, depending on the decomposition of supersymmetry, are dual
to one another.  In particular in this way we find the dual for the
$N=4$ models of \cite{DiKaVa}, corresponding to the
decomposition $4=0+4$, as $SU(2)$ holonomy compactifications
corresponding to the decomposition $4=2+2$.

We shall first discuss examples with $N=6,4$ and $2$.
To proceed further to other values of $N$ we have
to go beyond the adiabatic argument.
 In order to do this we follow an idea,
though not justifiable by adiabatic argument, which we find reasonable.
This idea will be explained at the end of this section and some examples
with $N=1,2$ supersymmetries will be constructed along that line.

\noindent{\bf N=6 Example:}
We choose,
\be \label{e66}
(\theta_L,\phi_L,\theta_R,\phi_R)=(\pi,-\pi, 0, 0) .
\ee
Eq.\refb{e23} now gives,
\be \label{e67}
(\theta_L',\phi_L',\theta_R',\phi_R')=(\pi, -\pi, 0, 0) \, .
\ee
Note that the holonomies in the two theories are the same. Also, the
lattice with this symmetry is the original lattice $\Gamma_0^{(4,4)}$.
This gives $\langle M \rangle = \langle M'\rangle =I_8$.
Thus the model is self-dual.

The associated shifts $v$, $v'$
along one of the circles (say the fourth direction)
by half the periodicity is represented by the lattice vector
\be \label{e68}
v = v' = \pmatrix{\ha \cr 0\cr 0\cr 0\cr}
\ee
on $\Gamma_0^{(2,2)}$. The adiabatic
argument by itself does not rule out the possibility of adding a
vector $(0,0,\ha,0)$ to $v$ and /or
$v'$, but the left-right level matching condition determines
both $v$ and $v'$ uniquely to be given by eq.\refb{e68}.

The transformation $\Omega$ given in eq.\refb{e66} corresponds to
an $SU(2)$ holonomy on the left-movers and trivial
holonomy on the right-movers, giving us an $N=6$ model.
The self-duality of the model
reflects the fact that the SO(5,5;Z)
element $\wb\Omega_0$ commutes with the $T$-duality transformation
generated by $\Omega$, and hence is an element of the
U-duality group of the resulting theory. The spectrum of massless states
in this theory can be found as follows. First we note that under the $Z_2$
subgroup of the SO(4,4;Z) group generated by \refb{e66}, the different
SO(4,4) representations decompose as,
\ben \label{ek1}
8_v &=& 4(+) \oplus 4(-) \nonumber \\
8_s &=& 4(+) \oplus 4(-) \nonumber \\
8_c &=& 4(+) \oplus 4(-) \, .
\een
Thus we get four vector fields from $A^\ab_\bmu$
($0\le \bmu\le 3$), four from $K^\alpp_\bmu$,
four (=$4\times 2/2$) from the field strengths $D^\alp_{i\bmu\bnu}$
($4\le i\le 5$)
and four more vector fields from $G_{i\bmu}$
and $B_{i\bmu}$
(since $D$ is self-dual, two $D^\alp_{i\bmu\bnu}$ make one
unconstrained vector field).
This gives a total of 16 vectors which is the right number for the
N=6 supergravity theory in four dimensions\cite{CrJu}.
Similar counting shows that we
get the right number (30) of scalar fields as well.

As noted in\cite{VaWi2} the $U$-duality
group for $N=4,5,6$ will depend on the precise choice
of compactification.  In this example, part of the T-duality group
associated with the torus $T^2$ is given by $\Gamma_0(2)_T\times
\Gamma_0(2)_U$. Since this model is self-dual, it has an S-T exchange
symmetry which we shall call $(Z_2)_{S-T}$. This also shows that the
S-duality group is $\Gamma_0(2)_S$. The full U-duality group $G$ can be found
following the procedure indicated in the previous section.

The moduli space of the theory locally has the structure SO$^*$(12)/U(6).
Thus the global structure of the moduli space is given by
$G\backslash SO^*(12)/U(6)$.

\noindent {\bf N=4 Example A:}
 We now turn to constructing $N=4$ dual pairs.  We choose,
\be \label{e33}
(\theta_L,\phi_L,\theta_R,\phi_R)=(2\pi, 0, 0, 0) .
\ee
Eq.\refb{e23} now gives,
\be \label{e34}
(\theta_L',\phi_L',\theta_R',\phi_R')=(\pi, -\pi, \pi, -\pi) \, .
\ee
The transformation $\Omega$ given in eq.\refb{e33} corresponds to
$(-1)^{F_L}$ since it involves a rotation by $2\pi$ in the $6_L-7_L$
plane. On the other hand, the transformation $\Omega'$ corresponds to
an inversion of $T^4$, {\it i.e.} an inversion of the lattice
$\Gamma^{(4,4)\prime}$.
The associated shifts ($v$, $v'$)
along one of the circles (say the fourth direction)
by half the periodicity is represented by the lattice vector
\be \label{e35}
v = v' = \pmatrix{\ha \cr 0\cr 0\cr 0\cr}
\ee
on $\Gamma_0^{(2,2)}$.

We now turn to a comparison of the two theories.

\begin{itemize}

\item{ In the first theory (obtained after modding out by $\Omega$
together with the lattice translation $v$) the space-time supersymmetry
generators on the left are all broken, whereas those on the right
are all intact. Thus this theory has
$(0_L,4_R)$ supersymmetry in four dimensions.
This gives a total of N=4 supersymmetry.
On the other hand, the second theory has $(2_L,2_R)$ space-time supersymmetry,
which again gives N=4 supersymmetry in four dimensions.
Note that the $(2_L,2_R)$ theory is a geometric compactification
of strings on an $SU(2)$ holonomy manifold which is
not $K3\times T^2$.
}
\item{ In the first theory all states that come from the Ramond sector
on the left are projected out. This, in particular, means that there
are no states from the RR sector. Thus the gauge fields that survive
the projection are $A^\ab_\bmu$, $G_{i\bmu}$ and $B_{i\bmu}$.
This gives a
total of 12 gauge fields. On the other hand, in the second theory,
the gauge fields $A^{\prime\ab}_\bmu$, which originate from the internal
components $G^{[10]}_{m\bmu}$, $B^{[10]}_{m\bmu}$ ($6\le m\le 9$), as well
as the gauge fields which originate from
the components $C^{[10]}_{mi\bmu}$,
get projected out since they are odd under the inversion of the
four torus, but the gauge fields $K^{'\alpp}_\bmu$, $G'_{i\bmu}$ and
$B'_{i\bmu}$ survive. This again gives a total of 12 gauge fields.
}

\item{ Since in the $N=4$ supergravity theories the structure of the
low energy effective action is completely determined by the number
of U(1) gauge fields, the two theories have identical low energy effective
action. In particular, the local structure of the moduli space in both the
theories is given by
$\big(SL(2,R)/U(1)\big) \times \big( O(6,6)/O(6)\times O(6)\big)$.
In the first theory the $SL(2,R)/U(1)$ component of the moduli space
is parametrized by the axion-dilaton field $S$, whereas the moduli
$T$ and $U$ of $T^2$, the internal moduli of $T^4$, and the Wilson lines
of the NS sector gauge fields parametrize the $\big( O(6,6)/
O(6)\times O(6)\big)$ component of the moduli space. On the other
hand, in the second theory the $SL(2,R)/U(1)$ component of the
moduli space is parametrized by the modulus $T'$ of $T^2$. The
axion-dilaton field $S'$, the modulus $U'$ of $T^2$, the internal
moduli of $T^4$, and the RR Wilson lines parametrize the coset
$\big(O(6,6)/O(6)\times O(6)\big)$. This is consistent with our
previous assertion that the field $S$ should be identified with
$T'$, and that $T$ should be identified with $S'$.
}

\end{itemize}

Using the duality between these two theories we can get information about the
duality symmetry group of the individual theories. In particular,
according to the argument given before, in both the theories the
duality groups acting on the S, T and U fields
are given by $\Gamma_0(2)$. Thus globally
the moduli space has the structure
\be \label{e42}
\big(\Gamma_0(2)\backslash
SL(2,R)/U(1)\big)\times \big(G\backslash O(6,6)/O(6)\times
O(6)\big)\, ,
\ee
where
\be \label{e43}
G\cap O(2,2)=\Gamma_0(2)\times \Gamma_0(2)\, .
\ee

Finally we turn to the study of enhanced gauge symmetries.
In the first theory, by adjusting $G_{ij}$, $B_{ij}$ ({\it i.e.}
the moduli $T$ and $U$), or equivalently, by performing an O(2,2) boost with
the matrix
\be \label{ematrix}
\V^{-1}={1\over\sqrt 2} \pmatrix{0 & 0 & 0 & 1 \cr 2 & 0 & 0 & 1 \cr
2 & 2 & -1 & 1 \cr 0 & 0 & 1 & -1 \cr} \, ,
\ee
we can bring the lattice $\Gamma_0^{(2,2)}$ to the
form $\Gamma^{(2,2)}(D_2)$ where:
\ben \label{e39}
\Gamma^{(l,l)}(D_{l}) = \Bigg\{ {1\over \sqrt 2}
\pmatrix{m_1 \cr \cdot \cr \cdot \cr m_l \cr n_1 \cr \cdot \cr \cdot
\cr n_l \cr} \Bigg\}
\, , \quad & &  m_i, n_i \in Z, \quad \sum_{i=1}^l m_i , \, \sum_{i=1}^l n_i
\,\, \in 2Z\, , \nonumber \\
& & m_i + n_i + m_j + n_j \, \in \, 2Z \,\, \forall \, \, (i,j) \, .
\nonumber \\
\een
The same boost brings the shift vector $v$ to the form:
\be \label{e40}
v = {1\over\sqrt 2} \pmatrix{1 \cr 0 \cr 1\cr 0\cr}\, ,
\ee
up to a lattice translation. According to the analysis of ref.\cite{DiKaVa}
this theory has an enhanced gauge symmetry $(SU(2))^2$.\footnote{Note
that we are using the basis in which the metric is $\pmatrix{0 & I_2\cr
I_2 & 0 \cr}$ instead of $\pmatrix{-I_2 & 0 \cr 0 & I_2}$ used in
ref.\cite{DiKaVa}.}
Then the equivalence of the two theories implies that the second
theory also has an enhanced gauge symmetry for these values of the moduli.
The map between the fields in the two theories shows that the massless
charged vector particles in the second theory carry magnetic type
$B'_{i\bmu}$ charge and hence are non-perturbative states in the spectrum.
Furthermore, in order to reach the enhanced symmetry point, we need to
adjust the moduli $U'$ and $S'$ in the second theory. Thus the symmetry
enhancement is a non-perturbative phenomenon in the second theory.

We can get further symmetry enhancement in the first theory by
adjusting the moduli of $T^4$ and $T^2$, as well as the
Wilson lines associated with the gauge field components $A^\ab_i$.
This is equivalent to an O(6,6) boost. With the help of this
boost we can bring the lattice to the form $\Gamma^{(6,6)}(D_6)$
and the shift vector to the form
\be \label{e40a}
v = {1\over\sqrt 2} \pmatrix{1 \cr 0^5 \cr 1\cr 0^5\cr}\, ,
\ee
This gives an enhanced symmetry group $SU(2)^6$
(studied in\cite{BlDoGo,Kaw,Les,FeKo,DiKaVa}). In the second theory,
in order to
reach these enhanced symmetry points we need to switch on background
values of the fields $K^{\prime\alpp}_i$. These correspond to RR
Wilson lines in the theory.
Also some of the
massless charged vector bosons carry charge under the U(1) gauge fields
$K^{\prime\alpp}_\bmu$ which originate in the RR sector.
Thus we see again that the mechanism of symmetry
enhancement is a non-perturbative effect in the second theory.

\noindent {\bf N=4 Example B}: In this set of examples we choose
\be \label{e44}
(\theta_L, \phi_L, \theta_R, \phi_R) = ({4\pi\over n}, 0, 0, 0)\, ,
\qquad n\ge 3.
\ee
This gives
\be \label{e45}
(\theta'_L, \phi'_L, \theta'_R, \phi'_R) = ({2\pi\over n}, -{2\pi\over n},
{2\pi\over n}, -{2\pi\over n})\, .
\ee
Physically, $\Omega$ given in eq.\refb{e44} corresponds to a rotation
by $4\pi/n$ in the $(6_L,7_L)$ plane, whereas $\Omega'$ given in
eq.\refb{e45} corresponds to a left-right symmetric rotation by
$2\pi/n$ in the $(6,7)$ plane, and by $-2\pi/n$ in the $(8,9)$ plane.
For both the theories we choose the lattice shift in $\Gamma_0^{(2,2)}$
to be
\be \label{e46}
v = v' = \pmatrix{1/n \cr 0 \cr 0 \cr 0 \cr}\, .
\ee
As in previous examples, adiabatic argument does not rule out the possibility
of adding a term proportional to $(0,0,1/n,0)$ to $v$
and / or $v'$, but the requirement of left-right level matching prevents the
addition of such a vector. As in example A, both the theories have
N=4 supersymmetry in four dimensions, with the first theory having
$(0_L,4_R)$ supersymmetry, and the second theory having $(2_L,2_R)$
supersymmetry.

The analysis of the spectrum of massless U(1) gauge fields can be carried
out as follows. In the first theory, all the fields coming from the
RR sector are projected out, so we only need to concentrate on the NS
sector fields. Under the U(1) subgroup of SO(4,4) generated by
$(\theta,0,0,0)$ the vector representation of SO(4,4) decomposes as
\be \label{e48}
8_v = (+1) \oplus (-1) \oplus 6 (0) \, .
\ee
Thus of the eight vector fields $A^\ab_\bmu$, the six that are neutral under
this U(1) subgroup survives the projection. Together with the four
vector fields coming from the components $G_{i\bmu}$, $B_{i\bmu}$, these
constitute ten U(1) vector fields.

In the second theory, the relevant U(1) subgroup is generated by
$(\theta, -\theta, \theta, -\theta)$. The various SO(4,4) representations
have the following decomposition under this subgroup:
\ben \label{e50}
8_v & = & 4 (+1) \oplus 4 (-1)\, , \nonumber \\
8_c & = & (+2) \oplus (-2) \oplus 6(0) \, , \nonumber \\
8_s & = & 4 (+1) \oplus 4(-1) \, .
\een
Thus of the possible vector fields $A^{'\ab}_\bmu$, $K^{\prime\alpp}_\bmu$,
and those associated with the field strengths $D^{\prime\alp}_{i\bmu\bnu}$,
only six from $K^{\prime\alpp}_\bmu$ are
neutral under this SO(2), and hence survives this projection. Together
with the gauge field components $G'_{i\bmu}$, $B'_{i\bmu}$, this gives
ten U(1) gauge fields altogether, in agreement with the calculation in the
first theory.

As argued before, in this case the duality groups acting on the S, T and
U fields in both
theories are given by $\Gamma_0(n)$. Thus the moduli space has the
structure
\be \label{e52}
\big( \Gamma_0(n)\backslash SL(2,R) / U(1)\big) \times \big( G \backslash
O(6,4) / O(6) \times O(4) \big) \, ,
\ee
where
\be \label{e53}
G\cap O(2,2) = \Gamma_0(n) \times \Gamma_0(n) \, .
\ee

The choice of the holonomy $\Omega$ is the same as
that appears in some of the examples\cite{DiKaVa}, however
the choice of the lattice we have made is different from that chosen
there.  In particular the lattices studied there, which do lead
to enhanced gauge symmetries for $n=3$ corresponding to $SO(5)\times SU(3)$
and $SU(3)\times SU(3)$, for $n=4$ corresponding to $SU(4)\times SU(2)$
and for $n=6$ corresponding to $SO(5)\times SU(2)\times SU(2)$ ,
do not have a decomposition as $\Gamma^{(4,4)}+\Gamma^{(2,2)}$.
However by turning on gauge field components $A^\ab_i$ ($i=4,5$),
and choosing appropriate
shift vectors in the internal theory which correspond to
turning on RR Wilson lines in the dual theory we can obtain all these
models as special cases of the above dual pairs.\footnote{Another
example considered in\cite{DiKaVa} with gauge group $G_2$
is likely to be dual to the non-geometric $(2_L,2_R)$ compactification
with left-right
holonomy $SU(2)$ manifold for which we mod out by
$(\pi,-\pi,0,0)$ accompanied by the adiabatic shift $(1/2,0,0,0)$
and by $(0,0,\pi, -\pi)$ accompanied by the adiabatic shift $(0,1/2,0,0)$.}

\noindent {\bf N=2 Example A}: We shall now give an example of a self-dual
N=2 compactification. We choose,
\be \label{e55}
(\theta_L,  \phi_L, \theta_R, \phi_R) = ( 2\pi, 0, \pi, -\pi)\, .
\ee
This gives
\be \label{e56}
(\theta'_L,  \phi'_L, \theta'_R, \phi'_R) = ( 2\pi, 0, \pi, -\pi)\, .
\ee
Also, the lattice with this symmetry is the original lattice
$\Gamma_0^{(4,4)}$. This corresponds to $\langle M \rangle = \langle
M'\rangle=I_8$.

Physically, both $\Omega$ and $\Omega'$ represent the transformation
$(-1)^{F_L}$ together with an inversion of the right hand part of the
lattice $\Gamma^{(4,4)}$ associated with $T^4$. We accompany this
transformation with a lattice shift on $\Gamma_0^{(2,2)}$ of the form:
\be \label{e57}
v = v' = \pmatrix{{1\over 2} \cr 0 \cr 0 \cr 0 \cr}\, .
\ee
Since the two theories are identical, we have here an example of a self-dual
theory. Since under the duality transformation $S$ and $T$ gets exchanged,
we see that this theory is invariant under the exchange $S\leftrightarrow T$.

The supersymmetry is completely broken in the left sector, and half broken
in the right sector. Thus this theory has $(0_L, 2_R)$, {\it i.e.}
N=2 space-time supersymmetry. To calculate the massless spectrum in the
theory, we study the decomposition of various representations of the
SO(4,4) group under the SO(2) subgroup generated by $(2\theta, 0, \theta,
-\theta)$:
\ben \label{e59}
8_v & = & (+2) \oplus (-2) \oplus 2 (0) \oplus 2 (+1) \oplus 2(-1) \, ,
\nonumber \\
8_c & = & (+2) \oplus (-2) \oplus 2(0) \oplus 2(+1) \oplus 2(-1) \, ,
\nonumber \\
8_s & = & 2(+1) \oplus 2(-1) \oplus (+2) \oplus (-2) \oplus 2(0) \, .
\een
Since $\Omega$ corresponds to SO(2) rotation by $\pi$, we need to look
for states that carry even SO(2) charge. Thus we get four gauge fields from
$A^\ab_\bmu$, four from $K^\alpp_\bmu$,
four (=$4\times 2/2$) field strengths from $D^\alp_{i\bmu\bnu}$
and four more gauge fields from $G_{i\bmu}$ and $B_{i\bmu}$.
This gives a total of 16 vector fields, which correspond
to 15 vector multiplets.

Counting of scalars proceeds in a similar manner. First of all, the condition
that the lattice associated with $T^4$ is invariant under inversion on the
right completely fixes $\langle M \rangle$,
so we do not get any scalar moduli fields
from there. We get four scalars from $\psi^\alp$, eight (=$4\times 2$) from
$A^\alp_i$, eight (=$4\times 2$)
from $K^\alpp_i$, four from the field strengths $D^\alp_{ij\bmu}$, four
from $G_{ij}$, $B_{ij}$, and two from the dilaton-axion combination.
This gives a total of 30 scalar fields, which form part of the 15 vector
multiplets. Thus the theory does not contain any massless hypermultiplet.

The twisted sector in this theory for some values of the scalar moduli
(for example the one corresponding to the moduli where in the
$N=4$ example A gave $SU(2)^2$ enhanced gauge symmetry)
will contain charged massless hypermultiplets.  Thus we expect
that the moduli space for vector multiplets receives quantum corrections.
This is in accord with the fact that for this model the dilaton
is in the vector multiplet.

\noindent {\bf N=2 Example B}:
If we consider a $ Z_2\times Z_2$ orbifold consisting of
the modding outs we used in the $N=4$ example A and $N=2$ example
A, where we use the adiabatic shift vectors to be on two
different circles of $\Gamma_0^{(2,2)}$ we obtain another $N=2$
model.  Now, however the model is not self-dual.
The model and its dual receive their supersymmetry by $(0_L,2_R)$
and thus have the dilaton in the vector multiplet. It is easy
to see that generically the model has 7 vector multiplets and
no hypermultiplets and that the dilaton is in a vector multiplet
(compared to the previous example all the vector multiplets coming
from the RR sector have been projected out by the second $Z_2$).
For some values of the scalar moduli,
for example the one corresponding to the moduli where in the
$N=4$ example A gave $SU(2)^2$ enhanced gauge symmetry,
this theory has perturbative non-abelian $SU(2)$ gauge symmetry
with no matter.  We thus expect there to be quantum corrections
to the moduli geometry in accord with the fact that the dilaton
is in a vector multiplet.

\noindent {\bf N=2 Example C}:
Another example of a self-dual N=2 model is obtained by modding out the
theory by a $Z_2\times Z_2$ group generated by
\ben \label{es1}
\Omega_1 = (2\pi, 0, 0, 0)\, , & & \qquad v_1 = \pmatrix{\ha \cr 0 \cr
0\cr 0\cr}\, , \nonumber \\
\Omega_2 = (\pi, -\pi, \pi, -\pi)\, , & & \qquad v_2 = \pmatrix{0 \cr \ha \cr
0 \cr 0\cr}\, ,
\een
The duality map exchanges $\Omega_1$ and $\Omega_2$. Thus the theory is
self-dual provided we exchange the two circles of $T^2$. This
corresponds to the exchange $S\leftrightarrow T$, $U\to -1/U$.
At a generic point in the moduli space the theory has three vector
multiplets and four hypermultiplets.
Perturbatively, enhanced SU(2) gauge symmetry occurs at points in the
moduli space where the $N=4$ example A develops an enhanced $SU(2)^2$
gauge symmetry,
but the structure of the moduli space near this point
is corrected acoording to the results of Seiberg and Witten\cite{SeWi}.
(The perturbative version of such corrections have been worked out in
refs.\cite{KLL,AFGNT}.) The $S\leftrightarrow T$ symmetry of the theory
means that a similar structure of the theory must appear for large $T$
as well.

\noindent {\bf Beyond Adiabatic Argument}:
Even though the adiabatic argument seems to be powerful enough
to suggest many new dual pairs, it has its limitation as many
interesting cases and in particular models with $N=1$ supersymmetry cannot
be constructed in this way.  In this section we would like
to abstract a general lesson from the validity of the adiabatic
argument and suggest other ways of constructing dual pairs.
This will open up our hands in constructing a much larger
class of dual pairs.

One of the key checks of having a dual pair is the perturbative agreement
of the massless spectra of the two models at generic points in the moduli
space (special points may arise where perturbatively massless
states in one model will have solitonic interpretation in the dual model).
The adiabatic method of constructing dual pair already satisfies this
for the following reason:  We start with the massless modes in the higher
dimensional theory (in our case $d=6$) and take their momenta to be
independent of the directions we are compactifying (in our case
further compactification on $T^2$).  Moreover we keep only the
massless state invariant under the automorphism to be modded
out by.  Clearly the massless states we thus keep in the two
theories are in one to one correspondence since we have an explicit isomorphism
that relates the two theories and the corresponding automorphisms.
Moreover the fact that the twisted sector is generically massive
(as follows from the fact that we are going a fraction around a circle)
implies that we have already accounted for all the massless states, and
thus the two theories agree as far as the massless modes are concerned.

Note that
from the point of view of the low dimensional observer,
the states coming from the twisted sector are massive due to
the Bogmol'nyi bound for the mass.  It is thus natural to extend this
principle also to the contribution to the Bogomol'nyi bound from the
central charges arising from the {\it internal} $T^4$.  In other
words, if we mod out by transformations, all of which have
shift vectors that generically have components along the central charges,
then the corresponding twisted sector
states are generically all massive and thus the previous
argument for agreement between the massless states for the two theories
will continue to hold.  Using this idea we can construct dual
models with any value of $N$ (excluding $N=7$).
 We will now illustrate
this idea by constructing dual models with $N=1$ and $N=2$ supersymmetry
in four dimensions as well as an $N=2$ model in six dimensions.

\noindent {\bf N=2 Example D}:
We consider a $Z_2\times Z_2$ orbifold where we start from the
$Z_2$ orbifold model of the $N=4$ example A.  We mod out further
by a $Z_2$ transformation which acts on the internal $T^4$ by
\be \label{e88}
(\theta_L',\phi_L',\theta_R',\phi_R')=(\pi, 0, \pi , 0) .
\ee
accompanied by a left-right symmetric half-vector shift on both $\theta$
and $\phi$ planes
and a reflection on the common $T^2$.  The dual to this second $Z_2$
according to Eq.\refb{e23} is given by,
\be \label{e89}
(\theta_L,\phi_L,\theta_R,\phi_R)=(\pi, 0, \pi,0) \, ,
\ee
accompanied by RR gauge transformations corresponding
to the shifts we have introduced.
This corresponds to turning on half units of flux of some of the RR
gauge fields at the fixed points of this $Z_2$ transformation on $T^4$.
It is easy to see that
this theory has $N=2$ supersymmetry where in the primed
theory it comes from $(1_L,1_R)$ and in the unprimed
theory it comes from $(0_L,2_R)$.  Note also that in the unprimed
theory the fact that we get nothing from the twisted sector is
very much analogous to a similar situation considered in
ref.\cite{ScSe2}.
This model has generically
3 vector multiplets and 4 hypermultiplets.  As far as vector moduli
is concerned we can use the primed theory (where the dilaton is in the
hypermultiplet) to compute the exact moduli space, and we
see that it has moduli space $(\Gamma_0(2)\backslash SL(2,R)/U(1))^3$,
describing the $T$-moduli associated with the $\theta'$-plane, the
$\phi'$-plane, and the torus $T^2$.
This agrees with the classical
moduli space for the unprimed theory. Thus there is no quantum correction
in accord with the fact that there in the vector moduli there is
never an enhanced gauge symmetry point.  The hypermultiplet moduli
can be computed from the unprimed theory which gives us the standard
moduli space $G\backslash O(4,4)/O(4)\times O(4)$ describing the
four-torus involving the $\theta$-plane and $T^2$. The discrete duality
group $G$ is a subgroup of $SO(4,4;Z)$ which leaves the shift vector $v$
given in eq.\refb{e35} and the background RR flux invariant.
This again agrees with the classical moduli
of the primed theory.  Thus that receives no corrections as well.

\noindent {\bf N=1 Example}:
Let us now construct an $N=1$ example of dual pairs.
We consider a $Z_2\times Z_2$ orbifold.  One of the $Z_2$'s
is again the one considered in the self-dual $N=2$ Example A.
The second $Z_2$ we consider is the same one considered
in the last example, i.e. given by
\be \label{e887}
(\theta_L',\phi_L',\theta_R',\phi_R')=(\pi, 0, \pi , 0) .
\ee
accompanied by a left-right symmetric half-vector shift on both $\theta$
and $\phi$ planes
and a reflection on the common $T^2$. This has the same holonomy dual
but with RR fields turned on.  This theory has $N=1$ supersymmetry
coming from $(0_L,1_R)$.  Generically the theory has 9 chiral
multiplets (including the dilaton) and 6 vector multiplets
(four of the $U(1)$ gauge fields come from RR sector).
The gauge symmetry is always abelian, consistent with the fact
that we do not expect gaugino condensation with supersymmetry breaking.

{\bf 6-dimensional Examples:} Continuing in the same spirit, we can
construct dual pairs of models in six dimensions. Consider, for example,
the $N=2$ supersymmetric theory in six dimensions obtained by modding
out the maximally supersymmetric ($N=4$) theory by
$(\theta_L,\phi_L,\theta_R,\phi_R)=(2\pi, 0, 0, 0)$, together with a
left-right symmetric half shift in the $\phi$ plane. This gives a
$(0_L,2_R)$ theory with 8 U(1) gauge fields. Its dual is given by
modding out the same theory by $(\theta_L',\phi_L',\theta_R', \phi_R')=
(\pi, -\pi, \pi, -\pi)$, {\it i.e.} an inversion on the torus, together
with a RR gauge transformation. This introduces a flux of RR gauge
fields at the orbifold points and makes the twisted sector states
massive. Thus we now get an $N=2$ supersymmetric ($1_L,1_R$) theory with
8 U(1) gauge fields. Construction of this dual pair is very similar in
spirit to the construction of the dual of the CHL string discussed in
ref.\cite{ScSe2}. As in \cite{ScSe2}, we expect this theory to have a
solitonic string with half the tension of a fundamental string.

We would like to acknowledge the hospitality of the Aspen
Center for Physics. We would like to thank C. Hull and N. Warner for
discussion.  The research of C. Vafa was partially supported by
NSF grant PHY-92-18167. A. Sen would also like to acknowledge the
hospitality of Queen Mary and Westfield College and The University of
Wales, Swansea during the course of the work.

\end{document}